\documentclass[12pt,twoside]{article}
\usepackage{amsfonts}

\makeatletter
\newcommand{\lb}{\linebreak}
\newcommand{\bls}[1]{\advance\baselineskip -#1pt}

\newcommand{\starting}{\topmargin=-2.5cm
\renewcommand{\@evenfoot}{\roman{page} \hfil}
\renewcommand{\@oddfoot}{\hfil \roman{page}}}
\newcommand{\heads}[2]{%
\pagestyle{myheadings}
\renewcommand{\@oddhead}{\raisebox{0pt}[\headheight][0pt]{%
   \vbox{\hbox to\textwidth{\rightmark \hfil {\small\thepage} \strut}\hrule}}}
\renewcommand{\@evenhead}{\raisebox{0pt}[\headheight][0pt]{%
   \vbox{\hbox to\textwidth{{\small\thepage} \hfil \leftmark \strut}\hrule}}}
\markboth{\protect\small\it #1}{\protect\small\it #2}}
\newdimen{\Ione}
\newdimen{\Ipoltora}
\def\Title#1#2#3{%
\baselineskip=\Ipoltora
\begin{center}
{\Large\bf
\uppercase{#1} \\ }
\bigskip\bigskip
{#2} \\
{#3} \\
\end{center}}
\long\def\Abstract#1{%
\bigskip
\parbox{0.93\textwidth}{%
\begin{center}
{\bf Abstract} \\
\end{center}

\medskip
{
\baselineskip=\Ione

#1

}
\vss
}
\bigskip
}
\def\@cite#1#2{{$^{\hbox{\scriptsize#1})}$\if@tempswa , (see #2)\fi}}
\def\thebibliography#1{\section*{\refname\@mkboth
 {\uppercase{\refname}}{\uppercase{\refname}}}\list
 {\arabic{enumi}}{\settowidth\labelwidth{[#1]}\leftmargin\labelwidth
 \advance\leftmargin\labelsep
 \usecounter{enumi}}
 \def\newblock{\hskip .11em plus .33em minus .07em}
 \sloppy\clubpenalty4000\widowpenalty4000
 \sfcode`\.=1000\relax}
\makeatother

\def\be{\begin{equation}}
\def\ee{\end{equation}}
\def\p{\partial}
\def\lb#1{\label{#1} }
\def\d{{\rm d}}
\def\dfrac#1#2{{\displaystyle\frac{#1}{#2}}}
\def\mg{{\mathfrak g}}
\def\i{{\rm i}}
\def\dsum#1#2{{\displaystyle\sum\limits^{#1}_{#2}}}
\def\bea{\begin{eqnarray}\samepage}
\def\eea{\end{eqnarray}}

\def\stTD#1#2{\hbox to 0em{\mathsurround=0em %
$\stackrel{#1}{#2}$\hss} \phantom{#2}}
\def\stscript#1#2{\hbox to 0em{\mathsurround=0em %
${\scriptstyle\stackrel{#1}{#2}}$\hss} \phantom{#2}}
\def\stscriptscript#1#2{\hbox to 0em{\mathsurround=0em %
${\scriptscriptstyle\stackrel{#1}{#2}}$\hss} \phantom{#2}}
\def\st#1#2{\mathchoice{\stTD{#1}{#2}}{\stTD{#1}{#2}}%
{\stscript{#1}{#2}}{\stscriptscript{#1}{#2}}}

\begin{document}


\Title{Induced gravitation\\ as nonlinear electrodynamics effect}
{\large Alexander A. Chernitskii}
{St.-Petersburg Electrotechnical University\\aa@cher.etu.spb.ru}

%

\Abstract%
    {The effect of induced Riemann geometry in nonlinear electrodynamics
is considered. The possibility for description of real gravitation
by this effect is discussed.}

\section{Introduction}

The idea about a possibility for description of gravitation as
manifestation of another interactions was said by Sakharov\cite{Saharov}
in context of quantum field theory. Adler\cite{SLAdler1980}
proposed to consider the gravitation as manifestation of electromagnetic
interaction again in the scope of quantum field technique.
Such view to gravitation is in the direction of unified approach for
all interactions of material objects.

But in classical nonlinear electrodynamics, in particular, in Born-Infeld model
there is an effect which can be considered as means for description of
gravitation. This effect can be named as induced Riemann geometry
in nonlinear electrodynamics. It appear when we consider the problem on
movement of electromagnetic soliton or propagation of light with some
given field, for example, of distant solitons. This problem can be considered
as initial part of iterative procedure for obtaining complex solution
including many solitons and radiation field. The given field modifies
the trajectories of soliton and light beam. This modification can
appear as effective Riemann space for the movement of solitons and the
propagation of light, with metric depending on the given electromagnetic field
components.

For the first time, seemingly,
the effective Riemann space appearing in nonlinear electrodynamics
was discovered by \mbox{Plebansky\cite{Plebansky1968}} in the problem on
propagation of light in given field for Born-Infeld model.
Effective Riemann space in the problem for movement of electromagnetic
soliton in given field was discovered by
\mbox{Chernitskii\cite{AACher1992}}
for Mie nonlinear electrodynamics and for Born-Infeld type
models (Chernitskii\cite{AACher1993}).

At the present time Born-Infeld electrodynamics
(Born \& Infeld\cite{BornInfeld})
 arouses considerable interest
from various points of view. In particular, it appear also in
quantized strings theory (Fradkin \& Tseytlin\cite{FradkinAndTseytlin1985}).
Here we shall consider just Born-Infeld nonlinear electrodynamics model.

\section{Born-Infeld electrodynamics}

The Born-Infeld electrodynamics
is derived from a variational principle proposed by
Eddington\cite{Eddington1924}. He also had in view some unified approach.
This principle has the following very geometrical form:
\be
\lb{VarPr}
\delta\int\sqrt{|\det\left(g_{\mu\nu} {}+{} \chi\,F_{\mu\nu}\right)|}\,(\d x)^4
{}={} 0
\quad,
\ee
where the Greek indices take values \mbox{$0,1,2,3$},
\mbox{$g_{\mu\nu}$} are components of metric for space-time,
\mbox{$F_{\mu\nu}$} are components of the tensor of electromagnetic field,
\mbox{$\chi$} is a dimensional constant.

\section{Effective Riemann space}

The Born-Infeld system of equations
(see also Chernitskii\cite{Chernitskii1998a})
has the following very
notable form of the characteristic equation (Chernitskii\cite{I1998JHEP}):
\bea
\label{Eq:Char}
&&{\mg}^{\mu\nu}\,\frac{\partial \Phi}{\partial x^\mu}\,
\frac{\partial \Phi}{\partial x^\nu} {}={} 0
\quad,
\\
&&{\mg}^{\mu\nu} {}\equiv{} g^{\mu\nu} {}-{} \chi^2\,T^{\mu\nu}
\quad,
\label{Def:EffMetr}
\eea
where \mbox{$T^{\mu\nu}$} are components of symmetrical energy-momentum
tensor,
\mbox{$\Phi (x) {}={} 0$} is a characteristic surface.

It should be noted that form (\ref{Eq:Char}) of the characteristic equation
is typical for a broad class enough of nonlinear electrodynamics models
(see Glinskii\cite{Glinskii}, Novello et al\cite{Novellogrqc9911085}).
But for the Born-Infeld
electrodynamics the tensor \mbox{$\mg^{\mu\nu}$} includes
 (\ref{Def:EffMetr}) the energy-momentum tensor of the model!

As it is known, if we have some given field
\mbox{$\stackrel{g}{F}_{\sigma\rho}$},
the problem for propagation of quick-oscillating
wave with small amplitude gives the dispersion relation
coinciding with the characteristic equation (\ref{Eq:Char}) where
\mbox{$\Phi (x)$} is wave phase and
\mbox{${\mg}^{\mu\nu} {}={} {\mg}^{\mu\nu}
(\stackrel{g}{F}_{\sigma\rho})$}. Thus the light propagates
in the effective Riemann space induced by the given electromagnetic field.

The matter is constructed by charged particles which are in
permanent movement. Thus a massive body can look as a very complicated
non-stationary electromagnetic field configuration. We can use
Fourier expansion for this field configuration in proper coordinate system
\mbox{$\{y^\mu\}$} of the body. We assume here that in the proper
coordinate system this field configuration is time-periodical
but this is not essential and the period can rush to infinity.
Thus for moving body in the coordinate system \mbox{$\{x^\nu\}$}
we can write the following:
\be
\lb{Sol:BidyonFourier}
{\cal D}
{}={} \dsum{\infty}{f {}={} -\infty}
{\cal D}_f\,
\exp\left(\i f \Theta\right)
\quad,
\ee
where \mbox{${\cal D}$}  is the column with six components
of the electromagnetic field,
\mbox{${\cal D}_f {}={} {\cal D}_f (y^i)$},\\
\phantom{where}
the Latin indices (except $f$) take values $1,2,3$,\\
\phantom{where}
\mbox{$y^i {}={} L^i_j\left[x^j {}-{} a^j(x^0)\right]$}\quad,\qquad
\mbox{$L^\mu_\nu$} are component of Lorentz transformation matrix,\\
\phantom{where}
\mbox{$a^j {}={} V^j\,x^0 {}+{} a^j_0$}\quad
is the rectilinear trajectory,\\
\phantom{where}
\mbox{$\underline{\omega}\,y^0 {}\equiv{} \Theta {}={} \Theta (x)$}\quad,\qquad
\mbox{$\dfrac{\p \Theta }{\p x^\mu} {}\equiv{} k_\mu$}\quad,\qquad
and
\be
\lb{DispRelLin}
|g^{\mu\nu}\,k_\mu\,k_\nu| {}={} \underline{\omega}^2
\quad.
\ee

The static field configuration \mbox{${\cal D}_0$} in (\ref{Sol:BidyonFourier})
is the aggregate of distributed charges such that the full charge equals zero.
The movement of charged electromagnetic solitons in small given field
\mbox{$\stackrel{g}{F}_{\sigma\rho}$}
was considered by Chernitskii\cite{AACher1993}
for Born-Infeld type electrodynamical models.
Here the given field
is small with respect to the maximum value for the field
of the soliton under consideration.
Using an approximate method we obtain the trajectory \mbox{$a^j(x^0)$} modified by the small given
field.
The first order by the small field \mbox{$\stackrel{g}{F}_{\sigma\rho}$}
gives the Lorentz force (see also Chernitskii\cite{Idyonint}).
The second order gives the trajectory in the form
of geodesic line equation for some effective Riemann space
with metric depending on squares of the field components
\mbox{$\stackrel{g}{F}_{\sigma\rho}$}. Thus sign of charge does not
matter for this effect.

The quick-oscillating part of the field configuration
(\ref{Sol:BidyonFourier})
(for \mbox{$f {}\neq{} 0$}) is some standing wave in proper coordinate
system. The simplest example for such standing wave is the function
\be
\lb{Sol:StandWave}
\dfrac{\sin(\underline{\omega}\, r)}{\underline{\omega}\,r}\,
\sin(\underline{\omega}\,y^0)
\quad,\qquad
r {}={} \sqrt{y^i\,y_i}
\quad,
\ee
which is the solution for wave equation in spherical coordinate system.
If we operate on this function by the Lorentz transformation or
consider it in the coordinate system \mbox{$\{x^\mu\}$}
we obtain the prototype for the quick-oscillating part of
(\ref{Sol:BidyonFourier}).

The movement of such semi-standing wave having a small amplitude was
considered
by Chernitskii\cite{Idyonint}.  The additional given field
\mbox{$\stackrel{g}{F}_{\sigma\rho}$} modifies the originally
rectilinear trajectory of the semi-standing wave such that the dispersion
relation (\ref{DispRelLin}) is modified to the following:
\be
\lb{DispRelMod}
|\mg^{\mu\nu}\,k_\mu\,k_\nu| {}={} \underline{\omega}^2
\quad,
\ee
where \mbox{$\mg^{\mu\nu}$} are defined by (\ref{Def:EffMetr}) with
\mbox{${\mg}^{\mu\nu} {}={} {\mg}^{\mu\nu}
(\stackrel{g}{F}_{\sigma\rho})$}.

Relation (\ref{DispRelMod}) is coincide with the Hamilton-Jacobi
equation for a massive particle moving in gravitational field.
For the case \mbox{$\underline{\omega} {}={} 0$}
we have the dispersion relation for light.

The stated results argue for that the massive bodies, like the light,
 will be moving
into the effective Riemann space induced by the electromagnetic
field of distant bodies.

\section{Gravitation}

Let us write the following zero speed approximation for the geodesic
line equation of the effective Riemann space:
\be
\dfrac{\d V_i}{\d x^0} {}={} -\dfrac{1}{2}\,
\dfrac{\p {\mg}_{00}}{\p x^i}
\quad.
\lb{Eq:Newton}
\ee
For sufficiently small given field we can write
\be
{\mathfrak g}_{00} {}={} 1   {}-{} 2\,\epsilon\quad,
\qquad\mbox{where}
\qquad
\epsilon {}\equiv{} \dfrac{\chi^2}{4}\,
\left(\st{g}{{\bf E}}^2  {}+{} \st{g}{{\bf B}}^2\right)
\quad.
\lb{EffffMetr}
\ee

Let us define the Newtonian potential $\varphi$ as averaging of
\mbox{$\epsilon$} on a small four-dimensional volume \mbox{$\delta X^4$}
for removal of quick-oscillating part. Thus we have
\be
\varphi {}={} -\dfrac{1}{\delta X^4}
\int\limits_{\delta X^4}\epsilon\; (\d x)^4
\quad.
\lb{Def:Newton1}
\ee
On the other hand in Newtonian gravitational theory we have
\be
\varphi {}={} -G\,\dfrac{M}{r}
\quad.
\ee

Let us define the given field as the sum of two quick-oscillating
parts: field of massive neutral body at essentially distant region
from the body \mbox{$\stackrel{\circ}{F}_{\mu\nu}$}
and some background radiation \mbox{$\stackrel{\approx}{F}_{\mu\nu}$}.
We can write the following relations:
\be
\stackrel{\circ}{F}_{\mu\nu} {}\sim{}
\dfrac{\sin(\underline{\omega}\, r {}+{}
\stackrel{\circ}{\phi}_{r})}{\underline{\omega}\,r}\,
\sin(\underline{\omega}\,t {}+{} \stackrel{\circ}{\phi}_{t})
\quad,\qquad
\stackrel{\approx}{F}_{\mu\nu} {}\sim{}
\sin({\bf k}\cdot {\bf x} {}+{} \stackrel{\approx}{\phi}_{r})\,
\sin(\underline{\omega}\,t {}+{} \stackrel{\approx}{\phi}_{t})
\quad.
\lb{xxx}
\ee
It should be noted that  there is an interconsistency
between the phases of the fields
\mbox{$\stackrel{\circ}{F}_{\mu\nu}$} and
\mbox{$\stackrel{\approx}{F}_{\mu\nu}$} because of interaction. Thus from
(\ref{Def:Newton1}), (\ref{EffffMetr}), (\ref{xxx})
we can write the following evaluative expression:
\be
\varphi {}={} -\dfrac{\chi^2}{2}\,\left(\st{\circ}{G}\,M\,\dfrac{1}{r}\right)
\stackrel{\approx}{G}  {}+{} \,O\left(\dfrac{1}{r^2}\right)
\quad.
\ee
Here \mbox{$\st{\circ}{G}\,M/r$} characterizes the field of the
body \mbox{$\stackrel{\circ}{F}_{\mu\nu}$},
where we assume that it is proportional to the mass \mbox{$M$}.
And \mbox{$\stackrel{\approx}{G}$} characterizes the background field.
Thus we have
\be
G {}={} \dfrac{\chi^2}{2}\,\st{\circ}{G}\,\stackrel{\approx}{G}
\quad.
\ee

The model constant \mbox{$\chi$} can be obtained in electrodynamical
experiments (see, for example, Denisov\cite{Denisov2000}).

Most likely we can take \mbox{$\st{\circ}{G} {}={} {\rm const}$}. But
the background field can has weak space dependence, for example, on
density of distribution for the massive bodies and
\mbox{$\stackrel{\approx}{G} {}={} \stackrel{\approx}{G}(x)$}. Thus in this
approach the gravitational constant may be not constant. Perhaps,
the so called effect of dark matter can be explained with this argument.

\section{Conclusions}

Thus we have considered the effect of induced Riemann geometry in nonlinear
electrodynamics. This effect can account for real gravitation.

%

\end{document}